# High $T_c$ superconductivity at the interface between the $CaCuO_2$ and $SrTiO_3$ insulating oxides

D. Di Castro[1,2,*], C. Cantoni[3], F. Ridolfi[1], C. Aruta[2], A. Tebano[1,2], N. Yang[2,4], G. Balestrino[1,2]

[1]*Dipartimento di Ingegneria Civile e Ingegneria Informatica, Università di Roma Tor Vergata, Via del Politecnico 1, I-00133 Roma, Italy*

[2]*CNR-SPIN, Università di Roma Tor Vergata, Roma I-00133, Italy*

[3]*Materials Science and Technology Division, Oak Ridge National Laboratory, Oak Ridge, TN 37831-6116, USA*

[4]*Facoltà di Ingegneria, Università degli studi Niccolò Cusano, Rome I-00166, Italy*

*daniele.di.castro@uniroma2.it

**Abstract**

At interfaces between complex oxides it is possible to generate electronic systems with unusual electronic properties, which are not present in the isolated oxides. One important example is the appearance of superconductivity at the interface between insulating oxides, although, until now, with very low $T_c$. We report the occurrence of *high* $T_c$ superconductivity in the bilayer $CaCuO_2/SrTiO_3$, where both the constituent oxides are insulating. In order to obtain a superconducting state, the $CaCuO_2/SrTiO_3$ interface must be realized between the Ca plane of $CaCuO_2$ and the $TiO_2$ plane of $SrTiO_3$. Only in this case extra oxygen ions can be incorporated in the interface Ca plane, acting as apical oxygen for Cu and providing holes to the $CuO_2$ planes. A detailed hole doping spatial profile has been obtained by STEM/EELS at the O K-edge, clearly showing that the (super)conductivity is confined to about 1-2 $CaCuO_2$ unit cells close to the interface with $SrTiO_3$. The results obtained for the $CaCuO_2/SrTiO_3$ interface can be extended to multilayered high $T_c$ cuprates, contributing to explain the dependence of $T_c$ on the number of $CuO_2$ planes in these systems.

The interface between complex transition metal oxides is recently emerging as one of the most interesting systems in condensed matter physics [1]. Indeed, a range of fascinating interface electronic phenomena, as high mobility two-dimensional electron gas, quantum Hall effect, and superconductivity, have all been discovered in oxide hererostructures [2-8], but are absent in either of their constituents. One interesting example is the formation of a conducting quasi-two-dimensional electron gas at the interface between the insulating perovskites $LaAlO_3$ and $SrTiO_3$ [2]. Strikingly, this interface is also superconducting with transition temperature ($T_c$) about $10^{-1}$ K [4,9].

On the other hand, the layered structure of cuprate high $T_c$ superconductors (HTS) can be schematized as a sequence of natural interfaces between two blocks with different structure and functionality: an insulating block with "infinite layers" (IL) structure, i.e., containing a sequence of $CuO_2$ planes and Ca planes, and a charge reservoir (CR) block, that, opportunely doped by chemical substitution or excess oxygen ions, provides charge carriers to the IL block. The extraordinary electronic properties shown by the interface between two insulating oxides, as in LAO/STO heterostructures, suggested the opportunity to exploit the conducting interface as a charge reservoir to dope a cuprate with IL structure, and thus to dramatically raise the $T_c$ from ~ $10^{-1}$ K to ~ $10^2$ K. We recently explored such a possibility by tailoring artificial superlattices $[(CaCuO_2)_n/(SrTiO_3)_m]_N$, made by $N$ repetitions of the $(CaCuO_2)_n/(SrTiO_3)_m$ bilayer, where $n$ and $m$ are the number of unit cells of $CaCuO_2$ (CCO) and $SrTiO_3$ (STO), respectively. Indeed, it has been shown that the tetragonal phase of CCO, with $a = b = 3.86$ Å and $c = 3.20$ Å, can be stabilized in form of thin film by the good lattice match with perovkite substrates, in particular with $NdGaO_3$ [15]. The tetragonal phase shows an infinite layers structure, i.e., an infinite sequence of $CuO_2$ planes and Ca planes stacking along the c axis, as occurs in the IL blocks of HTS. Therefore, CCO is considered the parent compound of HTS with the simplest lattice structure. The $[(CaCuO_2)_n/(SrTiO_3)_m]_N$ superlattices are high $T_c$ superconductors with $T_c$ = 50 K [6, 10]. The overall experimental studies have suggested that the superconductivity is mainly confined at the interface between CCO and STO [6,10-12]. Differently from previous cuprate/cuprate heterostructures [5, 13, 14], now the spacer between the superconducting blocks is not a cuprate, but it is STO. The CCO/STO interfaces are thus very similar to the IL/CR native interface of HTS. The results obtained from the investigation of the CCO/STO interface can be used to extract important information on the physical processes occurring in HTS.

In this work, by using state-of-the-art aberration-corrected scanning transmission electron microscopy (STEM) coupled to electron energy loss spectroscopy (EELS), we investigate the CCO/STO interface present in the bilayer $(CCO)_n/(STO)_m$, with unprecedent spatial resolution [16]. We find that this interface shows high $T_c$ superconductivity, with optimal $T_c$ about 40 K, a value smaller, but still comparable with those found in the $[(CaCuO_2)_n/(SrTiO_3)_m]_N$ superlattices [6, 10]. We give *direct* evidence that charge doping, and therefore superconductivity, are realized through the introduction of extra oxygen ions at the CCO/STO interface, giving rise to a $CaO_x$ atomic plane interfaced with the $TiO_2$ plane of STO.

We used the pulsed laser deposition technique to synthesize the CCO/STO bilayers [16]. The films were deposited on NdO terminated 5x5 mm$^2$ NdGaO$_3$ (110) (NGO) oriented mono-crystalline substrates, the most suitable substrates to grow the CCO film [15] and CCO/STO superlattices [10]. The detailed structural characterization of the large number of CCO/STO heterostructures grown have been obtained by X-Ray Diffraction and STEM measurements [16].

As in CCO/STO superlattices [6], the conductivity, and thus the $T_c$, of CCO/STO bilayers vary substantially with varying the oxidizing power of the growth atmosphere, from an insulating to a superconducting behaviour, with maximum $T_c \approx 40$ K. On the other hand, in bare CCO films, only a slight variation of the resistance has been obtained by changing the oxidizing growth conditions, but no superconducting phase has been revealed (see also Ref. 6). This finding clearly indicates that the interface with the STO is the structural feature needed for the occurrence of superconductivity. However, since in the SL each CCO block is embedded between two STO blocks, two kinds of interface structures are present there, at the top and at the bottom side of the CCO: $CuO_2$-Ca-$TiO_2$-SrO and $TiO_2$-SrO-$CuO_2$-Ca, respectively, which could behave differently.

The bilayer CCO/STO on NdO-terminated substrate gives us the possibility to discriminate between the two kinds of interfaces (see Fig. 1). Indeed, if on the NdO-terminated NGO substrate the CCO is deposited first, resulting in the NGO/CCO/STO heterostructure, the CCO/STO interface presents the following sequence of atomic planes: $CuO_2$-Ca-$TiO_2$-SrO (see Fig.1b). Let us name this sequence interface α. On the contrary, if STO is deposited first, to form the heterostructure NGO/STO/CCO, the interface between STO and CCO is $TiO_2$-SrO-$CuO_2$-Ca (see Fig.1d). Let us call this stacking interface β. The two different atomic plane sequences have been confirmed by STEM, as shown below. In Fig.1a we report the R(T) curves for the two samples with different stacking order: only the NGO/CCO/STO system is superconducting. From these findings we deduce that, most probably, the interface capable to host superconductivity is $CuO_2$-Ca-$TiO_2$-SrO (see Fig.1b).

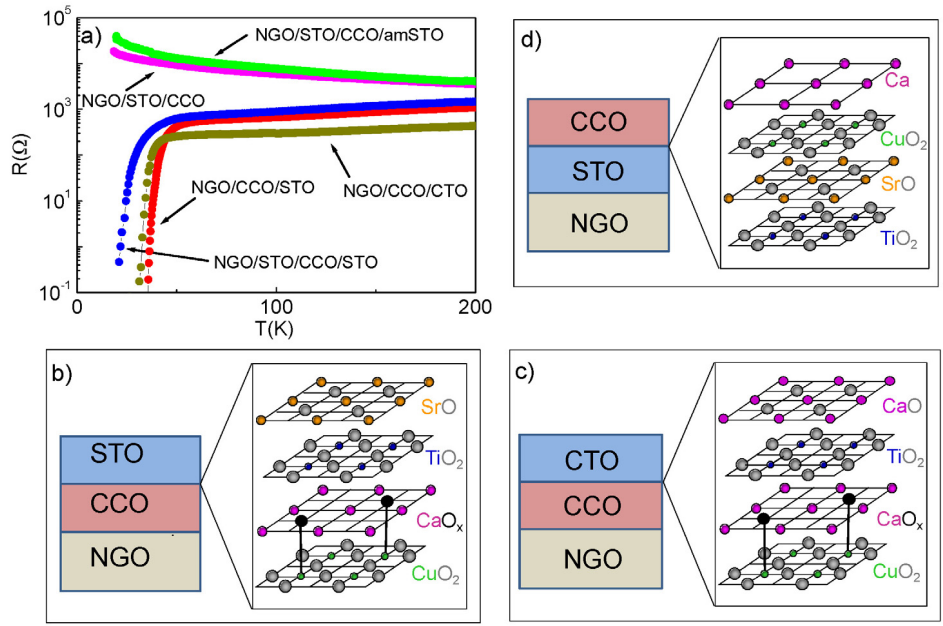

**Figure 1**: a) Resistance vs. temperature for the bilayer NGO/CCO/STO, the bilayer NGO/CCO/CTO, the bilayer NGO/STO/CCO, the trilayer NGO/STO/CCO/STO with the top crystalline STO layer, and the trilayer NGO/STO/CCO/amSTO with the top amorphous STO (amSTO) layer. The schematic interface structures for the bilayers NGO/CCO/STO (b), NGO/CCO/CTO (c) and NGO/STO/CCO (d) are shown. The extra oxygen ions introduced in the Ca plane at the interface in NGO/CCO/STO and NGO/CCO/CTO are shown in black instead of grey, and the corresponding apical coordination with the Cu ions is stressed by a thick black line.

This hypothesis has been further checked by growing a trilayer NGO/STO/CCO/STO. Here, both interfaces are again present and the superconducting behaviour is recovered (see Fig.1a). Moreover, if the top STO layer in the trilayer is made of amorphous STO (grown at low temperature), instead of crystalline epitaxial STO, the system remains insulating as for the NGO/STO/CCO bilayer (Fig.1a). This confirms that a sharp, crystalline $CuO_2$-Ca-$TiO_2$-SrO interface is necessary for superconductivity. Previous bulk-sensitive measurements on CCO/STO superlattices have suggested that charge doping results from extra oxygen atoms which are introduced in the system during the growth, most probably at the CCO/STO interfaces [6, 10, 11]. In agreement with this hypothesis, we note that, when the Ca plane is faced to the $TiO_2$ plane in the $CuO_2$-Ca-$TiO_2$-SrO interface, it can host extra oxygen ions giving rise to a $CaO_x$ plane, as indicated by the black dots in the interface scheme in Fig.1b. Every extra apical oxygen ion can inject two holes in the lower lying $CuO_2$ planes, thus leading to the occurrence of superconductivity. On the contrary, in the other interface, $TiO_2$-SrO-$CuO_2$-Ca, the SrO plane is already stoichiometrically full of oxygen ions (Fig.1d), and no hole doping can occur.

To experimentally check this hypothesis, we performed aberration-corrected STEM/EELS measurements on both kinds of interfaces at different points of the same samples and on different samples, obtaining similar results [16]. In Fig.2, we show STEM images of the CCO/STO interface of NGO/$(CCO)_9$/$(STO)_{55}$ bilayer grown on NdO terminated NGO substrate, where the $CuO_2$-Ca-$TiO_2$-SrO interface is present. The images are acquired simultaneously using both the annular bright field (ABF) detector, which is sensitive to scattering from light O atoms (panel a), and the high-angle annular dark field (HAADF) detector, which yields Z-contrast and is preferred for imaging the heavier cations (panel b) [16]. Because in ABF imaging the atoms appear dark on a light background, the ABF image in Fig. 2a has been inverted in order to show the atoms as bright spots on a dark background. In order to better visualize both cations and oxygen columns, an overlay of the HAADF image colored in red and the inverted ABF image colored in green is shown in Fig. 2c. In the overlay all the cations appear yellow/orange (overlay of red and green) while the O shows up in green. The intensity profiles of the inverted ABF image, taken at the first (panel d), second (panel e), and fourth (panel f) Ca planes (indicated by the pink, green, and blue arrows, respectively) indicate that a substantial amount of extra oxygen atoms are present only in the Ca plane at the interface with the $TiO_2$, thus forming a $CaO_x$ plane, whereas no extra oxygen is detected in the fourth Ca plane. This result reveals that, in CCO/STO, the charge reservoir role is played actually by a *single* atomic plane at the interface. The introduction of excess oxygen atoms only in the Ca plane of CCO at the interface with STO and not in the inner Ca planes can be explained by the presence of strain effect confined at the interface, together with the hybrid nature of the CCO/STO interface.

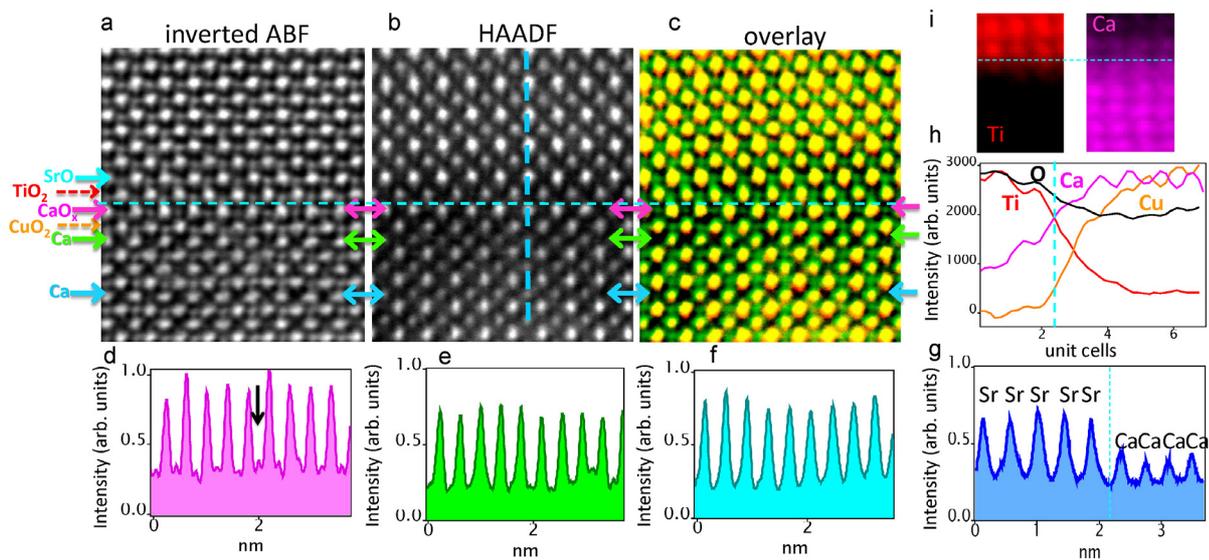

**Figure 2**: Structure of the CCO/STO interface in the NGO/CCO/STO bilayer: a) Inverted ABF images showing O columns. b) HAADF image showing cations with Z contrast. c) overlay of (a) in green and (b) in red, showing all the cations in orange (overlay of green and red) and the O columns in green. d), e), and f) are line profiles of the inverted ABF image at the magenta (interface Ca plane), green and blue arrow, respectively. h) EELS elemental profiles. g) intensity profile of the HAADF image (taken from the dashed blue line in b)). i) EELS elemental maps for Ti and Ca. Dashed cyan lines indicate the interface.

Indeed, depending on the oxygen content *x*, the interface $CaO_x$ plane can either couple with the $CuO_2$ plane below to form the infinite layer structure $CaCuO_2$, for the extreme case $x = 0$, or couple with the $TiO_2$ plane above to form the perovskite structure $CaTiO_3$, for the other extreme case $x = 1$, the latter being a stable compound in form of film. This opportunity makes the interface Ca plane unique with respect to the inner Ca planes of the CCO.

From the elemental maps and profiles obtained by EELS and shown in Fig. 2i,h the presence of some Ca/Sr interdiffusion at the interface is revealed [16]. However, the occurrence of Ca/Sr interdiffusion is not relevant for the appearance of superconductivity for two main reasons: 1) Ca is isovalent to Sr, therefore, no Sr doping can be envisaged; 2) the bilayer $CaCuO_2/CaTiO_3$, that is, with $CaTiO_3$ (CTO) instead of $SrTiO_3$, and thus with no possibility of Ca/Sr interdiffusion, is also superconducting with similar $T_c$, as shown in Fig.1a. In this case the interface is $CuO_2$-Ca-$TiO_2$-CaO (Fig.1c), with a structure identical to that in NGO/CCO/STO bilayer except for the presence of Ca instead of Sr in the last atomic plane. This last result is important since it generalizes the choice of the non-cuprate block from STO to many other possible insulating oxides. According to the above considerations about the IL/perovskite hybrid character of the CCO/STO interface, we suggest that, among all the possible $ABO_3$ (ABO) perovkites, the ones yielding superconducting CCO/ABO heterostructures are the ones for which the corresponding $CaBO_3$ compound exists in form of film.

Apart from the Ca diffusion in the STO, which can be also due to resputtering of light cations during ion-mill thinning, the CCO/STO interface is fairly sharp from the point of view of the other cations. Indeed, in the elemental profile (see Fig. 2h), Cu and Ti profiles decay much faster than the Ca profile, indicating that the $CuO_2$ plane below the interface is almost ideal. This fact can be considered a good condition for the occurrence of superconductivity. Moreover, the peak positions in the line profiles in panel h confirm the assumed $CuO_2$-Ca-$TiO_2$-SrO plane stacking.

The same measurements are shown in Fig.3 for the STO/CCO interface ($TiO_2$-SrO-$CuO_2$-Ca) taken on the trilayer NGO/STO/CCO/STO. Here, the intensity profiles show that in the interface SrO plane the oxygen ions are present (Fig.3f), as they should, in order to complete the neutral unit cell of STO ($TiO_2$-SrO). These oxygen ions do not dope the above lying $CuO_2$ planes with holes, since they are stoichiometric. In the Ca plane just above, no substantial number of excess oxygen ions is present (Fig.3d) and this interface is thus not conducting.

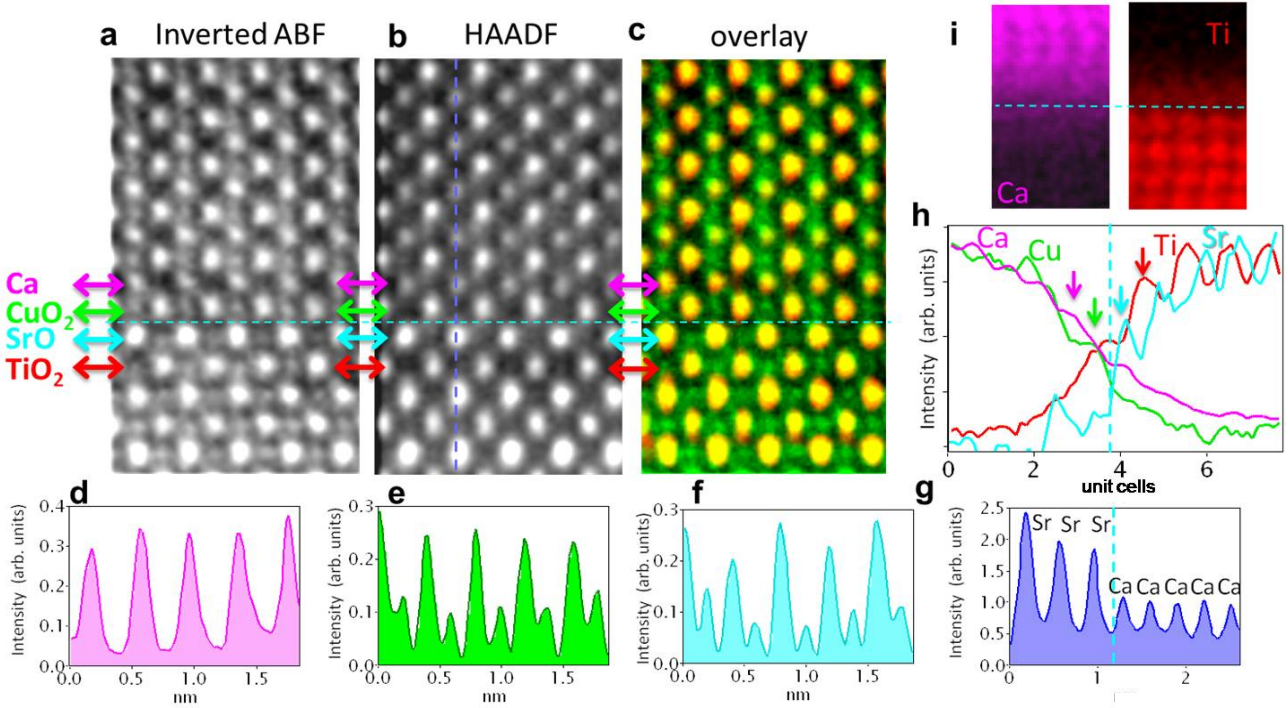

**Figure 3**: Structure of the STO/CCO interface in the NGO/STO/CCO/STO trilayer: a) Inverted ABF images. b) HAADF image. c) overlay of (a) in green and (b) in red, showing all the cations in orange and the O columns in green. d), e), and f) are line profiles of the inverted ABF image at the magenta (interface Ca plane), green (CuO2) and blue (SrO) arrows, respectively, showing no oxygen at the interface Ca plane. h). EELS elemental line profiles for Ca, Cu, Ti, and Sr confirming a quite sharp interface with stacking $TiO_2$-SrO-$CuO_2$-Ca. g) line profile taken from (b) at the blue line. i) EELS elemental maps for Ti and Ca. Dashed cyan lines indicate the interface position. Arrows in (h) indicate by color coding the stacking of atomic planes at the interface.

One may now wonder how far from the CCO/STO interface the holes, introduced by the extra oxygen ions, diffuse within the CCO layer. This, together with the interrelated dependence of $T_c$ on the number of $CuO_2$ planes, is also an open question in multilayered HTS [17-23].

Oxygen K-edge X-Ray Absorption spectroscopy has been demonstrated to be a powerful tool to reveal the presence of delocalized holes in HTS [24, 25]. Indeed, a pre-edge feature emerges upon doping HTS, which can be associated to the low-energy quasi-particle band called Zhang-Rice singlet: a locally bound $d^9$ copper 3d hole hybridized with a doped ligand hole (L) distributed on the planar oxygen 2p orbitals. The same feature can be observed in the EELS spectra at the O K-edge [26,27].

Thanks to the atomic spatial resolution of STEM/EELS, O-Kedge spectra could be acquired across the two types of interfaces with sub-angstrom steps. In the trilayer NGO/STO/CCO/STO, there are two interfaces: the interface with STO at the top (interface α), where extra oxygen ions are present (see Fig. 2), and the interface with STO at the bottom (interface β), where no extra oxygen ions are revealed (Fig.3). In panels a), b) and c) of Fig. 4 the O-K edge spectra acquired in the STO layer and in the CCO layers close to interface α and to interface β, respectively, are shown [16].

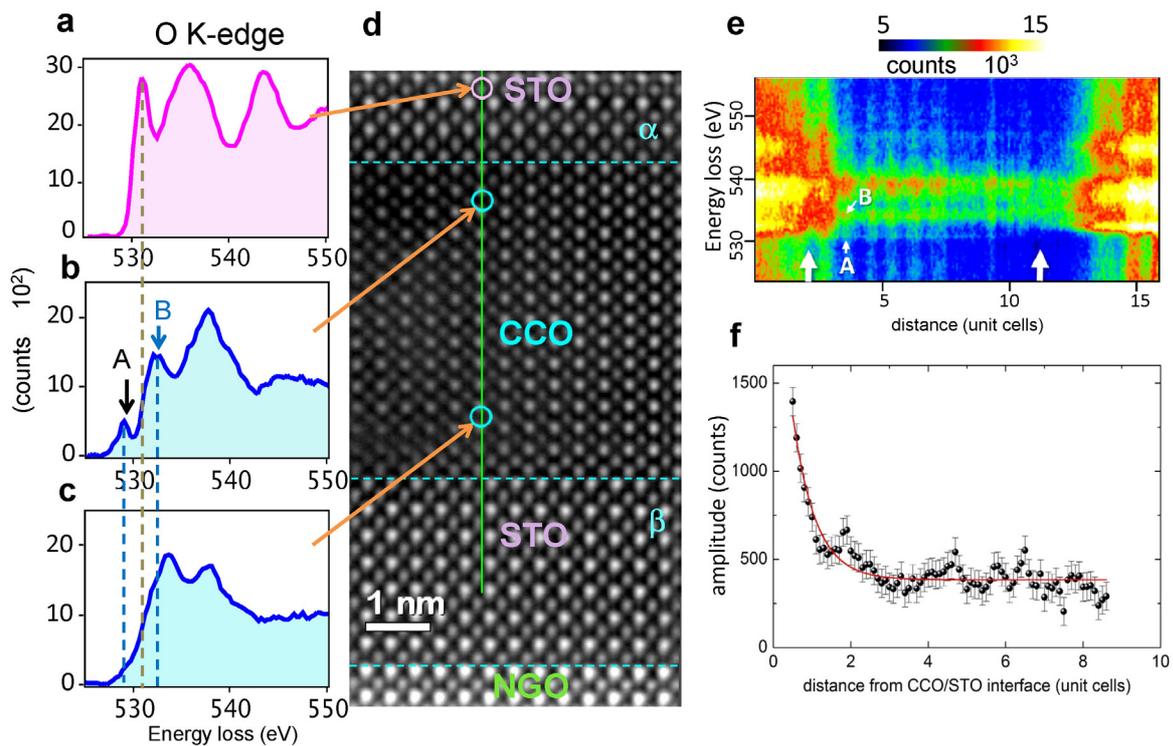

**Figure 4**: O K-edge measurements in the trilayer NGO/STO/CCO/STO: a) O K-edge from the upper STO layer in (d). b) O K-edge from the CCO rigth below the STO layer, corresponding to x = 1.2 u.c. in f. c) CCO O K-edge near the bottom STO layer, corresponding to x = 8.5 u.c. in f. Spectra a)-c) are shown after background subtraction and noise reduction by principal component analysis (PCA). d) HAADF image of the trilayer from which the spectrum image in (e) was taken. e) spectrum image (raw data) showing decreasing peak A amplitude in going from interface α to interface β (big white arrows indicate the range of data shown in f, small white arrows indicate the position of peak A and B as indicated in panel b). f) plot of the amplitude of A as a function of the distance *d* from the top CCO/STO interface. The line is a guide to the eyes.

In the vicinity of interface α, the spectrum of the CCO shows an additional spectral feature,

named A, around 529 eV. This peak disappears on going towards the other interface β and is absent in STO, which is an insulator.

According to previous XAS and EELS measurements [24-27], this spectral weight is ascribed to delocalized doping holes. Namely, peak A results mainly from $3d^9L \rightarrow 1s3d^9$ transitions, where L and 1s denote the O $2p_{x,y}$ ligand hole and O 1s core hole, respectively. Peak A is absent in the EELS spectrum of CCO taken far from the interface (Fig. 4c) and in the XAS and EELS spectra of all the insulating parents of high-$T_c$ cuprates [24-27]. In these compounds, the lowest energy peak, usually labeled as "B", occurrs at about 1-2 eV above peak A and is associated with the upper Hubbard band, specifically to the transitions from O 1s orbitals to O 2p orbitals hybridized with Cu 3d orbitals. In CCO, peak B is mostly masked by the leading edge of higher energy peaks and can only be resolved in high energy-resolution measurements after background removal [24]. Given the relatively low energy resolution of our EELS measurements (0.6 eV measured as FWM of the Zero Loss peak), the peak visible above A is in fact a convolution of the above mentioned upper Hubbard band peak ($3d^9 \rightarrow 1s3d^{10}$), and other stronger features originating from the overlap of Ca wavefunctions with O p-like states [16]. The dependence of the amplitude of peak A on the distance *d* from the top interface α is reported in panel f) of Fig. 4. The holes concentration decays rapidly with the distance from the top interface α and is drastically reduced already in the second CCO unit cell (second $CuO_2$ plane). This behavior, can be explained by the presence of an attractive electrostatic potential [20] associated with the negatively charged extra apical oxygen ions localized at the Ca interface plane.

We can consider the results shown in Fig.4 a *direct* measurement of the detailed dependence of the holes concentration on the number of $CuO_2$ planes in the IL block of HTS. From this point of view, the dependence of $T_c$ on the number *n* of $CuO_2$ planes in multilayered HTS [17-23], where the CR blocks inject carriers from both the top and the bottom interface (top and bottom interface of IL block with CR blocks are identical in HTS), can be explained. Given the dependence of the holes concentration on the distance from the interface shown in Fig. 4, the doping of the $CuO_2$ planes in HTS with $n \approx 3$ is homogeneous among the planes and the $T_c$ is maximum. For $n > 3$, the holes diffuse also to inner $CuO_2$ planes, the doping starts to be lower on average and not homogeneous, and the $T_c$ decreases. For $n \geq 4$-5, since the holes are confined by the electrostatic shielding within about 1-2 $CuO_2$ planes close to each interface of the IL block with the CR block, the inner planes are actually less than underdoped, and, most probably, antiferromagnetic, as also suggested in previous NMR studies [20]. Therefore, the increase of $n$ above 5 should not induce any change in $T_c$, which, indeed, keeps constant [23]. It is significant that for CCO/STO superlattices a $T_c$ vs. $n$ behaviour identical to the one found in HTS is reported [6]. This fact supports the reliability of the similitude between CCO/STO interface and IL/CR interface in HTS.

In conclusion, we have shown that the heterostructure CCO/STO with a single interface is superconducting at high $T_c$. The doping is achieved thanks to extra oxygen atoms, which, during the growth, are incorporated in the Ca plane of CCO at the interface with STO, giving rise to $CaO_x$ composition of the Ca interface atomic plane. A direct measurement of the concentration of the doping holes, introduced by the extra oxygen ions, as a function of the distance from the charge reservoir interface shows that they are confined within about 1-2 CCO unit cells from the interface. Deeper $CuO_2$ planes cannot be doped. The same behavior is expected to occur in standard multilayer HTS, where both the IL/CR interfaces are active, thus explaining the dependence of $T_c$ on the number of $CuO_2$ planes in these systems.

We conclude that the superconducting properties are mostly connected with the properties of the interface between the infinite layer cuprate and the non-cuprate block, such as termination, sharpness and capability of accommodating a large number of extra charges without diffusion of them toward the cuprate block. The non-cuprate block should be electronically inert to avoid charge shielding on the wrong side of the interface. The search for interfaces with improved superconducting properties can proceed by looking for other non-cuprate blocks having a better lattice match with CCO, a fully controlled termination and a better capability to accommodate extra oxygen ions.

The results reported in this work have also other important implications: they show that it is possible, in a relatively easy way, to isolate a single $CuO_2$ high $T_c$ superconducting plane. This is considered an important step towards the achievement of resistance-free electrical transport in 2D channels [29]. Indeed, this is connected to the possibility to realize new superconducting field effect devices, operating at temperatures much higher than the one realized with the LAO/STO heterostructure [9].

**Acknowledgement.**

This work was partially supported by Italian MIUR (PRIN Project 2010-2011 OXIDE, "OXide Interfaces: emerging new properties, multifunc-tionality, and Devices for Electronics and Energy". CC was supported by the U.S. Department of Energy, Office of Science, Basic Energy Sciences, Materials Sciences and Engineering Division.

# References

[1] H. Y. Hwang, *Mater. Res. Soc. Bull.* **31**, 28–35 (2006).

[2] A. H. Ohtomo, H.Y. Wang, *Nature* **427**, 423 (2004).

[3] A. Tsukazaki *et al., Science* **315,** 1388 (2007).

[4] N. Reyren *et al., Science* **317**, 1196-1199 (2007).

[5] A. Gozar *et al., Nature* **455,** 782 (2008).

[6] D. Di Castro *et al., Phys. Rev. B* **86**, 134524 (2012)

[7] J. E. Kleibeuker *et al., Phys. Rev. Lett.* **113**, 237402 (2014).

[8] R. Pentcheva *et al., Phys. Rev. Lett.* **104**, 166804 (2010).

[9] A. D. Caviglia *et al., Nature* **456**, 624 (2008).

[10] D. Di Castro *et al. Supercond. Sci. Technol.* **27,** 044016 (2014)

[11] C. Aruta *et al.*, *Phys. Rev. B* **87**, 155145 (2013)

[12] M. Salvato *et al. J. Phys.: Condens. Matter* **25,** 335702 (2013)

[13] G. Balestrino, S. Martellucci, P. G. Medaglia, A. Paoletti, and G. Petrocelli, *Phys. Rev. B* **58**, R8925 (1998)

[14] C. Aruta *et al., Phys. Rev. B* **78**, 205120 (2008).

[15] G. Balestrino *et al., J. Mater. Chem.* **5**, 1879 (1995).

[16] See Supplemental Material at [*URL will be inserted by publisher*] for details on experimental techniques and results.

[17] B. A. Scott *et al., Physica C* **230**, 239–245 (1994).

[18] Y. Tokunaga *et al., J. Low Temp. Phys*. **117,** 473 (1999).

[19] Y. Tokunaga *et al., Phys. Rev. B* **61,** 9707 (2000).

[20] H. Kotegawa, *J. Phys. Chem. Solids* **62**, 171 (2001).

[21] N. Hamada, H. Ihara, *Physica C* **357–360** 108 (2001).

[22]  S. Chakravarty, H-Y Kee, K. Voelker, *Nature* **428**, 53 (2004)

[23]  A. Iyo *et al., Physica C* **445 -448**, 17-22 (2006)

[24] T. Chen *et al., Phys. Rev. Lett*. **66**, 104 (1991)

[25] C. T. Chen, *Phys. Rev. Lett*. **68**, 2543 (1992).

[26] H. Romberg, M. Alexander, N. Nucker, P. Adelmann, and J. Fink, *Phys. Rev. B* **42**, 8768R (1990)

[27] N. Gauquelin *et al*., *Nature Communications* **5**, 4275 (2014)

[29] S. Gariglio, M. Gabay, J-M. Triscone, *Nature nanotechnology* **5**, 13 (2010)